\documentclass[aps,prd,preprint,nofootinbib]{revtex4}
\usepackage{epsfig}
\usepackage{graphicx}

\begin{document}
\title{The Color-Octet Contributions to $P$-wave $B_c$ Meson Hadroproduction\\[3mm]}
\author{Chao-Hsi Chang$^{1,2}$ \footnote{email:
zhangzx@itp.ac.cn}, Cong-Feng Qiao$^{1,4}$\footnote{email:
qiaocf@gscas.ac.cn}, Jian-Xiong Wang$^{3}$\footnote{email:
jxwang@mail.ihep.ac.cn}, Xing-Gang Wu$^{3}$\footnote{email:
wuxg@mail.ihep.ac.cn}}
\address{$^1$CCAST (World Laboratory), P.O.Box 8730, Beijing 100080,
China.\\
$^2$Institute of Theoretical Physics, Chinese Academy of Sciences,
P.O.Box 2735, Beijing 100080, China.\\
$^3$Institute of High Energy Physics, P.O.Box 918(4), Beijing 100049, China\\
$^4$Department of Physics, Graduate School of the Chinese Academy
of Sciences, Beijing 100049, China}
%\date{}

\begin{abstract}
The contributions from the color-octet components $|(c\bar b)_{\bf
8}(^{1}S_{0}) g\rangle$ and $|(c\bar b)_{\bf 8}(^{3}S_{1})
g\rangle$ to the $h_{B_c}$ or $\chi_{B_c}^J$ (the $P$-wave $B_c$
meson) hadroproduction are estimated in terms of the complete
${\cal O}(\alpha_s^4)$ calculation. As necessary inputs in the
estimate, we take the values of the octet matrix elements
according to the NRQCD scaling rules, and as a result, we have
found that the contributions to the $P$-wave production may be the
same in order of magnitude as those from the color-singlet ones,
$|(c\bar b)_{\bf 1}(^{1}P_{1})\rangle$ and $|(c\bar b)_{\bf
1}(^{3}P_{J})\rangle$ ($J=1,2,3$). Especially, our result
indicates that the observation of the color-octet contributions to
the $P$-wave production in the low transverse momentum region is
not very far from the present experimental capability
at Tevatron and LHC.\\

\noindent {\bf PACS numbers:} 12.38.Bx, 13.85.Ni, 14.40.Nd,
14.40.Lb.

\noindent {\bf Keywords:} inclusive hadroproduction, color-octet
components, $P$-wave states.
\end{abstract}

\maketitle

\section{Introduction}

$B_c$ meson has been observed experimentally \cite{CDF,D0}, and
within theoretical uncertainties and experimental errors the
observations are consistent with the theoretical predictions. In
view of the prospects in $B_c$ physics at the Fermilab Tevatron II
and LHC, $B_c$ physics is attracting more and more attentions. The
hadronic production of the $B_c$ meson (its ground state mainly),
has been studied quite well
\cite{prod,prod0,prod1,prod2,prod3,prod4,bcvegpy1,bcvegpy2}.
Especially, a generator BCVEGPY \cite{bcvegpy1,bcvegpy2} for
hadronic production of $B_c$ meson has been available, which can
be easily complimented to the PYTHIA environment \cite{pythia} and
enhances the efficiency to generate the full $B_c$ events greatly
in comparison with using PYTHIA itself.

In the framework of effective theory of NRQCD \cite{nrqcd,hqg}, a
heavy quarkonium is considered as an expansion of various Fock
states. The relative importance among those infinite ingredients
is evaluated by the velocity scaling rule. A similar idea can be
applied to the double heavy meson system too. That is, the
physical state of $B_c, B_c^*, h_{B_c}$ and $\chi^J_{B_c}$ mesons
can be decomposed into a series of Fock states:
\begin{eqnarray}
|B_c\rangle = {\cal O}(v^{0})|(c\bar{b})_{\bf 1}(^{1}S_{0})\rangle
+{\cal O}(v^1)|(c\bar{b})_{\bf 8}(^{1}P_{1})g\rangle +{\cal
O}(v^{1})|(c\bar{b})_{\bf 8}(^{3}S_{1})g\rangle +\cdots
\nonumber\\
|B_c^*\rangle = {\cal O}(v^{0})|(c\bar{b})_{\bf
1}(^{3}S_{1})\rangle +{\cal O}(v^1)|(c\bar{b})_{\bf
8}(^{3}P_{J})g\rangle +{\cal O} (v^{1})|(c\bar{b})_{\bf
8}(^{3}S_{1})g\rangle +\cdots \label{eq:1}
\end{eqnarray}
and
\begin{eqnarray}
|h_{B_c}\rangle = {\cal O}(v^{0})|(c\bar{b})_{\bf
1}(^{1}P_{1})\rangle +{\cal O}(v^1)|(c\bar{b})_{\bf
8}(^{1}S_{0})g\rangle +{\cal O}(v^{1})|(c\bar{b})_{\bf
8}(^{3}P_{J})g\rangle +\cdots
\nonumber\\
|\chi^J_{B_c}\rangle = {\cal O}(v^{0})|(c\bar{b})_{\bf
1}(^{3}P_{J})\rangle +{\cal O}(v^1)|(c\bar{b})_{\bf
8}(^{3}S_{1})g\rangle +{\cal O} (v^{1})|(c\bar{b})_{\bf
8}(^{1}P_{1})g\rangle +\cdots, \label{eq:2}
\end{eqnarray}
here $v$ is the relative velocity. Note that throughout this paper
we use the symbols $h_{B_c}$ and $\chi_{B_c}^J$ to denote the four
physical $P$-wave states ($h_{B_c}$ denotes the $P$-wave state
with the dominant color-singlet state $(c\bar{b})_{\bf 1}(^1P_1)$,
while $\chi_{B_c}^J$ denotes the $P$-wave states with the dominant
color-singlet states $(c\bar{b})_{\bf 1}(^3P_J)$ as described in
Eq.(\ref{eq:2})) instead of $B_{cJ,L=1}^*$ that is used in
Ref.\cite{cww}. Here the thickened subscripts of the $(c\bar{b})$
denote for color indices, ${\bf 1}$ for color singlet and ${\bf
8}$ for color-octet; the relevant angular momentum quantum numbers
are shown in the parentheses accordingly. According to the
velocity scaling rule, the probability of each Fock state in the
expansion is proportional to the scales in a definite power of
$v$, squared the indicated one as the above. Generally, the
leading Fock states of the $B_c$ and $B_c^*$ states are
$|(c\bar{b})_{\bf 1}(^1S_0)\rangle$ and $|(c\bar{b})_{\bf
1}(^3S_1)\rangle$ respectively, whose probability in respect of
the relative velocity are set to be the order of $O(v^0)$. Also
for the production of the $B_c$ and $B_c^*$ states, there is no
problem at all, one can be sure that the contributions from
leading Fock state(s) are dominant.

As implied in equations Eq.(\ref{eq:2}), at leading and next
leading order, the production of $P$-wave $B_c$ mesons may involve
in the Fock states: $|(c\bar b)_{\bf 1}(^{1}P_{1})\rangle$,
$|(c\bar b)_{\bf 8}(^{1}S_{0}) g\rangle$, $|(c\bar b)_{\bf
8}(^{3}P_{J}) g\rangle \cdots$; $|(c\bar b)_{\bf
1}(^{3}P_{J})\rangle$, $|(c\bar b)_{\bf 8}(^{3}S_{1}) g\rangle$,
$|(c\bar b)_{\bf 8}(^{1}P_{1}) g\rangle \cdots$. For color-singlet
components ($P$-wave), the long distance non-perturbative matrix
elements can be directly connected with the first derivative of
the wave functions at the origin, which can be computed via
potential models\cite{potential} and/or potential NRQCD
(pNRQCD)\cite{pnrqcd} and/or lattice QCD etc at the accuracy for
the present purposes, whereas, there is no reliable way to offer
such accurate values for the color-octet non-perturbative matrix
elements as those for color-singlet ones even for lattice QCD
\cite{hqg}. So far, we may only have rough values of these matrix
elements relating to the components $|(c\bar b)_{\bf 8}(^{1}S_{0})
g\rangle$, $|(c\bar b)_{\bf 8}(^{3}S_{1}) g\rangle \cdots$ by
means of the velocity power counting rule of the NRQCD effective
theory \cite{nrqcd}. In order to make a reasonable estimate on the
contributions from the color-octet components to the hadronic
production so as to see if the color-octet components are visible
experimentally or not, we take a similar way as done in
Ref.\cite{wccf}, i.e. the color-octet matrix elements are taken to
be smaller by certain order in $v^2$ than those obtained by
relating them to the color-singlet $S$-wave wavefunctions at the
origin $|\psi(0)|^2$. Whereas, the first derivative of the
wavefunctions ($P$-wave) at the origin for the color-singlet
components with proper factor of $m$ (the reduce mass of the bound
state), i.e. $\frac{\psi'(0)}{m}$, is in a higher order in $v$
than the wavefunctions at the origin for the $S$-wave functions,
$\psi(0)$, therefore the contributions from the color-octet
components $|(c\bar b)_{\bf 8}(^{1}S_{0}) g\rangle$, $|(c\bar
b)_{\bf 8}(^{3}S_{1}) g\rangle \cdots$ ($S$-wave) to the
production of the $P$-wave excited states $h_{B_c}$ and
$\chi^J_{B_c}$ may be comparable with those from the color-singlet
components $|(c\bar b)_{\bf 1}(^{1}P_{1})\rangle$ and $|(c\bar
b)_{\bf 1}(^{3}P_{J})\rangle$ ($P$-wave).

The color-singlet contributions to the $P$-wave $B_c$ meson
hadronic production have been studied in
Refs.\cite{bcvegpy2,cww,cheung,berezhnoy}. In Ref.\cite{cheung},
the hadronic productions of $h_{B_c}$ and $\chi_{B_c}^J$ were
estimated by applying the fragmentation approach. While in
Refs.\cite{bcvegpy2,cww,berezhnoy} the authors adopted the lowest
order (here $\alpha^4_s$) complete approach but only color-singlet
components $|(c\bar b)_{\bf 1}(^{1}P_{1})\rangle$ and $|(c\bar
b)_{\bf 1}(^{3}P_{J})\rangle$ were taken into account, and found
that the total cross-section is much larger than what predicted by
the fragmentation approach \cite{cheung}, and the fragmentation
results tend to be comparable with those from the $\alpha_s^4$
complete calculation only in the cases when the transverse
momentum $p_t$ of the meson $B_c$ is so high as $p_t\geq 30$ GeV
(higher than $S$-wave production case) \cite{berezhnoy,cww}. For
the $\alpha^4_s$ complete estimation, to be complete and to see
the characteristics, we compute the contributions from the
color-octet components to the $P$-wave $B_c$-meson production
precisely in the paper.

The paper is organized as follows: In Section II we will present
the basic formulae for the lowest order ($\alpha_s^4$) which are
used in the complete calculation of the contributions to the
$P$-wave hadroproduction from the color octet components $|(c\bar
b)_{\bf 8}(^{1}S_{0}) g\rangle $ and $|(c\bar b)_{\bf
8}(^{3}S_{1}) g\rangle$. In Section III, we will explain the
options about the input parameters, which are necessary for the
calculations and will present the numerical results properly. In
the last section, a brief summary and some discussions will be
given.

\section{Formulation and technique}

In hadron-hadron collisions at high energy, the gluon-gluon fusion
mechanism is the dominant one over the others. Hence, to study the
hadronic production of the $P$-wave $B_c$ mesons, as argued in the
preceding section, we need to consider two types of processes,
i.e., $gg\rightarrow (c\bar{b})_{\bf 1} +b+\bar{c}$ where
$(c\bar{b})_{\bf 1}$ is in color-singlet $^1P_1$ and $^3P_J$
states; and $gg\rightarrow (c\bar{b})_{\bf 8} +b+\bar{c}$ where
$(c\bar{b})_{\bf 8}$ is in color-octet $^1S_0$ and $^3S_1$ states.
In Ref.\cite{cww}, the color-singlet contributions to the hadronic
production of $h_{B_c}$ and $\chi_{B_c}^J$ have been computed, and
the useful formulas and techniques have been explained there. For
simplicity, we will not repeat it here. In the present work, our
main concern is to complete the calculations on the hadronic
production of the $P$-wave $B_c$ physical states, especially to
pay attention to the contributions from the color-octets $|(c\bar
b)_{\bf 8}(^{1}S_{0}) g\rangle$ and $|(c\bar b)_{\bf
8}(^{3}S_{1})g\rangle$.

The amplitude for the $gg\rightarrow (c\bar{b})_{\bf 8}+b+\bar{c}$
can be analytically obtained by applying the helicity method
\cite{bcvegpy1}, and $(c\bar{b})_{\bf 8}$ will be `hadronized' to
the state $|(c\bar b)_{\bf 8}(^{1}S_{0}) g\rangle$ or $|(c\bar
b)_{\bf 8}(^{3}S_{1}) g\rangle$ finally. The details of the
helicity technique can be found in Ref.\cite{bcvegpy1} and
references therein. The difference of the present calculations is
only in the color configuration from those for the leading order
$B_c (B_c^*)$ meson hadronic production.

In the case of the hadronic production $gg\rightarrow
(c\bar{b})_{\bf 1} +b+\bar{c}$ where $(c\bar{b})_{\bf 1}$ is in
color-singlet, $|(c\bar b)_{\bf 1}(^{1}S_{0})\rangle$ or $|(c\bar
b)_{\bf 1}(^{3}S_{1})\rangle$, there are only three independent
color factors (i.e. three independent color flows\cite{cww,cf}).
While for the case of the hadronic production of the color-octet
states, $gg\rightarrow (c\bar{b})_{\bf 8} +b+\bar{c}$ where
$(c\bar{b})_{\bf 8}$ is in color-octet and will be hadronized to
$|(c\bar b)_{\bf 8}(^{1}S_{0})g\rangle$ or $|(c\bar b)_{\bf
8}(^{3}S_{1})g\rangle$ finally, there are totally ten independent
color factors, $C_{kij}$ ($k=1,2,\cdots,10$), where $i,j=1,2,3$
are color indices of the quarks $\bar{c}$ and $b$ respectively.
They are
\begin{eqnarray}
C_{1ij}&=& \frac{1}{\sqrt{2}N_c}\left(T^bT^aT^d\right)_{ij}\;,\;\;
C_{2ij}=\frac{1}{\sqrt{2}N_c} \left(T^aT^bT^d\right)_{ij}\;,\nonumber\\
C_{3ij}&=& \frac{1}{\sqrt{2}}Tr\left[T^aT^d\right]T^b_{ij}\;,\;\;
C_{4ij}=\frac{1}{\sqrt{2}} Tr\left[T^bT^d\right]T^a_{ij}\;,\nonumber\\
C_{5ij}&=&
\frac{1}{\sqrt{2}}Tr\left[T^bT^aT^d\right]\delta_{ij}\;,\;\;
C_{6ij}=\frac{1}{\sqrt{2}}
Tr\left[T^aT^bT^d\right]\delta_{ij}\;,\nonumber\\
C_{7ij}&=& \frac{1}{\sqrt{2}N_c}\left(T^dT^bT^a\right)_{ij}\;,\;\;
C_{8ij}=\frac{1}{\sqrt{2}N_c}
\left(T^dT^aT^b\right)_{ij}\;,\nonumber\\
C_{9ij}&=& \frac{1}{\sqrt{2}N_c}\left(T^aT^dT^b\right)_{ij}\;,\;\;
C_{10ij}=\frac{1}{\sqrt{2}N_c} \left(T^bT^dT^a\right)_{ij}\;,
\end{eqnarray}
where the indices $a$ and $b$ are color indices for gluon-1 and
gluon-2 respectively, and $\sqrt{2} T^d$ stands for the color of
the color-octet state $(c\bar{b})_{\bf 8}$ in the production.
$N_c=3$, for QCD. All the color factors in the subprocess can be
expressed by the linear combination of the above ten independent
color factors, $i.e.,$ in the amplitude all the color factors may
be written in terms of these ten explicitly. Thus the total
helicity amplitude can be generically expressed as
\begin{eqnarray}
M^{(\lambda_{1}, \lambda_{4}, \lambda_{5},
\lambda_{6})}(q_{b1},q_{b2},q_{c1},q_{c2},k_{1},k_{2})
&=&\sum_{m=1}^{10}C_{mij}M^{(\lambda_{1},\lambda_{4},\lambda_{5},
\lambda_{6})}_{m}(q_{b1},q_{b2},q_{c1},q_{c2},k_{1},k_{2})\;,
\end{eqnarray}
where $\lambda_{1}$, $\lambda_{4}$, $\lambda_{5}$ and
$\lambda_{6}$ denote the helicities of the out-going $b$-quark,
out-going $\bar{c}$-quark, gluon-1 and gluon-2, respectively.
$k_{1}$ and $k_{2}$ are the momenta of the gluons; $q_{c1}$,
$q_{b1}$ are the momenta of $c$ and $b$ quarks and $q_{c2}$,
$q_{b2}$ are the momenta of $\bar{c}$ and $\bar{b}$ anti-quarks,
respectively. The helicity amplitude
$M^{(\lambda_{1},\lambda_{4},\lambda_{5},
\lambda_{6})}_{m}(q_{b1},q_{b2},q_{c1},q_{c2},k_{1},k_{2})$ can be
directly read from the Eqs.(4-8) in Ref.\cite{bcvegpy1},  with
only the color-singlet matrix elements (or the wave function at
the origin) being replaced by the color-octet matrix elements
$\langle 0|\chi_b^\dag T^d\psi_c (a_H^\dag a_H) \psi_c^\dag
T^d\chi_b |0\rangle$ and $\langle 0|\chi_b^\dag \sigma^i T^d
\psi_c (a_H^\dag a_H) \psi_c^\dag \sigma^i T^d \chi_b|0\rangle$.
To get the matrix element squared, one needs first to deal with
the square of the above ten independent color factors, i.e.
($C_{mij}\times C_{nij}^*$) with $m,n=1,2,\cdots 10$. For
reference use, whose values are listed in TAB.\ref{colortable}.

\begin{table}
\begin{center}
\caption{The square of the ten independent color factors
(including the cross terms), ($C_{mij}\times C_{nij}^*$) with
$m,n=1,2,\cdots,10$ respectively.}\vspace{3mm}
\begin{tabular}{|c||c|c|c|c|c|c|c|c|c|c|}
\hline ~ & ~$C_{1ij}^*$~ & ~$C_{2ij}^*$~ & ~$C_{3ij}^*$~ &
~$C_{4ij}^*$~ & ~$C_{5ij}^*$~ & ~$C_{6ij}^*$~ & ~$C_{7ij}^*$~ &
~$C_{8ij}^*$~& ~$C_{9ij}^*$~ & ~$C_{10ij}^*$~\\
\hline\hline ~$C_{1ij}$~ & $\frac{32}{81}$ & $-\frac{4}{81}$ &
$\frac{4}{9}$ & $-\frac{1}{18}$ & $\frac{7}{18}$ & $-\frac{1}{9}$
& $\frac{1}{162}$ & $\frac{10}{162}$ & $\frac{1}{162}$ & $-\frac{4}{81}$ \\
\hline ~$C_{2ij}$~ & $-\frac{4}{81}$ & $\frac{32}{81}$ &
$-\frac{1}{18}$ & $\frac{4}{9}$ & $-\frac{1}{9}$ & $\frac{7}{18}$
& $\frac{10}{162}$
& $\frac{1}{162}$ & $-\frac{4}{81}$ & $\frac{1}{162}$ \\
\hline ~$C_{3ij}$~ & $\frac{4}{9}$ & $-\frac{1}{18}$& $4$ &
$\frac{1}{2}$& $0$ & $0$ & $-\frac{1}{18}$& $\frac{4}{9}$&
 $\frac{4}{9}$& $\frac{4}{9}$ \\
\hline ~$C_{4ij}$~ & $-\frac{1}{18}$& $\frac{4}{9}$&
$\frac{1}{2}$& $4$ & $0$ & $0$ & $\frac{4}{9}$& $-\frac{1}{18}$ &
 $\frac{4}{9}$&$\frac{4}{9}$  \\
\hline ~$C_{5ij}$~ & $\frac{7}{18}$& $-\frac{1}{9}$& $0$ & $0$ &
$\frac{7}{2}$& $-1$ & $\frac{7}{18}$ & $-\frac{1}{9}$&
$\frac{7}{18}$ &$-\frac{1}{9}$  \\
\hline ~$C_{6ij}$~ & $-\frac{1}{9}$& $\frac{7}{18}$& $0$ &$0$
&$-1$ & $\frac{7}{2}$&$-\frac{1}{9}$ & $\frac{7}{18}$&
$-\frac{1}{9}$&  $\frac{7}{18}$\\
\hline ~$C_{7ij}$~ & $\frac{1}{162}$&
$\frac{10}{162}$&$-\frac{1}{18}$ &$\frac{4}{9}$ &
$\frac{7}{18}$&$-\frac{1}{9}$ & $\frac{32}{81}$&
$-\frac{4}{81}$& $\frac{1}{162}$& $-\frac{4}{81}$ \\
\hline ~$C_{8ij}$~ & $\frac{10}{162}$&$\frac{1}{162}$
&$\frac{4}{9}$ &$-\frac{1}{18}$ &$-\frac{1}{9}$ &$\frac{7}{18}$
&$-\frac{4}{81}$
& $\frac{32}{81}$& $-\frac{4}{81}$& $\frac{1}{162}$ \\
\hline ~$C_{9ij}$~ & $\frac{1}{162}$& $-\frac{4}{81}$&
$\frac{4}{9}$& $\frac{4}{9}$ & $\frac{7}{18}$&$-\frac{1}{9}$ &
$\frac{1}{162}$&
$-\frac{4}{81}$& $\frac{32}{81}$&$\frac{10}{162}$  \\
\hline ~$C_{10ij}$~ & $-\frac{4}{81}$&$\frac{1}{162}$ &
$\frac{4}{9}$ &$\frac{4}{9}$& $-\frac{1}{9}$&$\frac{7}{18}$ &
$-\frac{4}{81}$&
 $\frac{1}{162}$& $\frac{10}{162}$& $\frac{32}{81}$ \\
\hline
\end{tabular}
\label{colortable}
\end{center}
\end{table}

According to pQCD factorization theorem, the inclusive $B_c$ meson
hadroproduction can be formulated as
\begin{equation}
d\sigma=\sum_{ij}\int dx_{1}\int
dx_{2}F^{i}_{H_{1}}(x_{1},\mu^2_{F})\times
F^{j}_{H_{2}}(x_{2},\mu^2_{F})d\hat{\sigma}_{ij\rightarrow
(c\bar{b})X}(x_{1},x_{2},\mu^2_{F},\mu^2,Q^2)\,, \label{pqcdf}
\end{equation}
where $F^{i}_{H_1}(x,\mu^2_{F})\,,\;F^{j}_{H_2}(x,\mu^2_{F})$ are
parton distribution functions (PDFs) of partons $i$ and $j$ in
hadrons $H_1,\;H_2$, respectively. $\mu^2$ is the `energy scale
squared' where renormalization for the subprocess is made; $Q^2$
is the `characteristic energy scale of the subprocess squared';
and $\mu^2_F$ is the `energy scale squared' where the
factorization of the PDFs and the hard subprocess is made.
Usually, in the LO calculation, to avoid the large logarithms the
scales of typical scale of hard interaction, factorization and
`renormalization' are set to be the same, i.e.
$\mu^2=\mu_F^2=Q^2$.

\section{Numerical results}

For convenience and as a reference, in the numerical calculation
we take the values of the radial wave function at the origin and
the first derivative of the radial wave function at the origin as
those given by Refs.\cite{chen,potential}, i.e., $|R(0)|^2=1.54
GeV^3$ and $|R^{\prime}(0)|^2=0.201 GeV^5$ (namely the two values
roughly mean $v^2\sim 0.1$, if the reduced mass $m\sim 1.2 GeV$),
which relate to the non-perturbative matrix elements of the color
singlet production definitely. Whereas there is no reliable way to
compute the production color-octet matrix elements $\langle
0|\chi_b^\dag T^d\psi_c (a_H^\dag a_H) \psi_c^\dag T^d\chi_b
|0\rangle$ and $\langle 0|\chi_b^\dag \sigma^i T^d \psi_c
(a_H^\dag a_H) \psi_c^\dag \sigma^i T^d \chi_b|0\rangle$. Although
we do not know the exact values of the above two octet matrix
elements, according to NRQCD scale rule we know that they are at
the same order of the $P$-wave color-singlet matrix elements,
which are one order in $v^2$ higher than the $S$-wave
color-singlet matrix elements accordingly. Moreover according to
the heavy-quark spin symmetry\cite{nrqcd}, the values of the above
two production matrix elements have the approximate relation:
\begin{eqnarray}
\langle 0|\chi_b^\dag \sigma^i T^d \psi_c (a_H^\dag a_H)
\psi_c^\dag \sigma^i T^d \chi_b|0\rangle= (2 J + 1)\cdot\langle
0|\chi_b^\dag T^d\psi_c (a_H^\dag a_H) \psi_c^\dag T^d\chi_b
|0\rangle[1+{\cal O}(v^2)],
\end{eqnarray}
where $J$ is the total angular momentum of the hadron state. In
our numerical calculations, as done in Ref.\cite{wccf}, the values
of the color-octet matrix elements are set to be smaller by
certain order $v^2$ than those obtained by relating to the S-wave
wave functions at the origin $|\psi(0)|^2$ for the color singlet.
More specifically, based on the velocity scale rule, we can
estimate
\begin{eqnarray}
\langle 0|\chi_b^\dag T^d\psi_c (a_H^\dag a_H) \psi_c^\dag
T^d\chi_b |0\rangle &\simeq& \Delta_S(v)^2 \cdot \langle
0|\chi_b^\dag \psi_c (a_H^\dag a_H)
\psi_c^\dag \chi_b |0\rangle\nonumber\\
&\simeq& \Delta_S(v)^2\cdot \left|\langle 0 |\chi_b^\dag
\psi_c|B_c(^1S_0)\rangle\right|^2.\left[1+{\cal O}(v^4)\right]\,,
\end{eqnarray}
where the second equation comes from the vacuum-saturation
approximation. $\Delta_S(v)$ is of the order $v^2$ or so, and we
take it to be within the region of 0.10 to 0.30, which is in
consistent with the identification: $\Delta_S(v)\sim\alpha_s(Mv)$
and has covered the possible variation due to the different ways
to obtain the wave functions at the origin ($S$-wave) and the
first derivative of the wave functions at the origin ($P$-wave)
etc.

As argued in Ref.\cite{cww}, the equations: $P=q_{c1}+q_{b2}$,
$q_{c1}^2=m_c^2,\;q_{b2}^2=m_b^2$ and $P^2=M^2$ must be satisfied
simultaneously, and furthermore, only when these equations are
satisfied can the gauge invariance of the amplitude be guaranteed.
Hence, in this paper we proceed the numerical calculation by
making the choice: the masses of $c$ and $b$ quarks are $m_c=1.50$
GeV and $m_b=4.90$ GeV, respectively; the mass of the bound states
$|(c\bar b)_{\bf 1}(^{1}P_{1})\rangle$, $|(c\bar b)_{\bf
1}(^{3}P_{J})\rangle$, $|(c\bar b)_{\bf 8}(^{1}S_{0}) g\rangle$
and $|(c\bar b)_{\bf 8}(^{3}S_{1}) g\rangle$ are therefore to be
$M=6.40$ GeV, the sum of the two heavy quark masses.

There are a couple of uncertainty sources in the theoretical
estimates on the $B_c$ meson hadronic production, such as those
from the strong coupling constant, the $\alpha_s$; from the PDFs
of different sets of fittings {\it etc.}. In the present
calculations, the factorization energy scale is set to be the
transverse mass squared of the $(c\bar{b})$ bound states, i.e.,
$Q^2=M_t^2\equiv (M^2+p_{t}^2)$ where $p_t$ is the transverse
momentum of the bound state; the PDF of version CTEQ6L
\cite{6lcteq} and the leading order $\alpha_s$ running above
$\Lambda^{(n_f=4)}_{QCD}=0.326$ GeV are adopted.

\begin{table}
\begin{center}
\caption{Total cross-section (in unit of nb) for the hadronic
production of the $(c\bar{b})$ meson at LHC ($14.0$ TeV) and
TEVATRON ($1.96$ TeV), where for short the $|(^1S_0)_1\rangle$ denotes
$(c\bar{b})$ state in color-singlet $(^1S_0)$ configuration, and so forth.
Here $m_b=4.90$ GeV, $m_c=1.50$ GeV and $M=6.40$
GeV. For the color-octet matrix elements, we take
$\Delta_S(v)\in(0.10,\;0.30)$.} \vspace{2mm}
\begin{tabular}{|c||c|c||c|c||c|c|c|c|}
\hline\hline - &$|(^1S_0)_{\bf 1}\rangle$ & $|(^3S_1)_{\bf
1}\rangle$ & $|(^1S_0)_{\bf 8} g\rangle$ & $|(^3S_1)_{\bf 8}
g\rangle$ & $|(^1P_1)_{\bf 1}\rangle$ & $|(^3P_0)_{\bf 1}\rangle$
&$|(^3P_1)_{\bf 1}\rangle$ & $|(^3P_2)_{\bf 1}\rangle$ \\
\hline\hline LHC & 71.1 & 177. & (0.357, 3.21) & (1.58, 14.2) & 9.12 & 3.29 & 7.38 & 20.4\\
\hline TEVATRON & 5.50 & 13.4 & (0.0284, 0.256) & (0.129, 1.16) & 0.655 & 0.256 & 0.560 & 1.35\\
\hline\hline
\end{tabular}
\label{cross-sec}
\end{center}
\end{table}

In TABLE.\ref{cross-sec}, we show the total cross-sections for
hadronic production of the $(c\bar{b})$ meson at LHC and TEVATRON
respectively, where the $(c\bar{b})$ may mean in $|(c\bar b)_{\bf
1}(^{1}P_{1})\rangle$, $|(c\bar b)_{\bf 1}(^{3}P_{J})\rangle$,
$|(c\bar b)_{\bf 8}(^{1}S_{0}) g\rangle $ and $|(c\bar b)_{\bf
8}(^{3}S_{1}) g\rangle$ configurations respectively, and the
results of the $|(c\bar b)_{\bf 1}(^{1}S_{0})\rangle$ (the
dominant component for $B_c$) and $|(c\bar b)_{\bf
1}(^{3}S_{1})\rangle$ (the dominant component for $B_c^*$)
production are included for comparison. In the table, we take
$\Delta_S(v)\in(0.10,\;0.30)$. One may observe that the
cross-sections of the color-octet S-wave states are comparable to
those of the color-singlet $P$-wave states, and they are in the
same order in $v^2$ expansion as expected by NRQCD.

\begin{figure}
\centering
\includegraphics[width=0.460\textwidth]{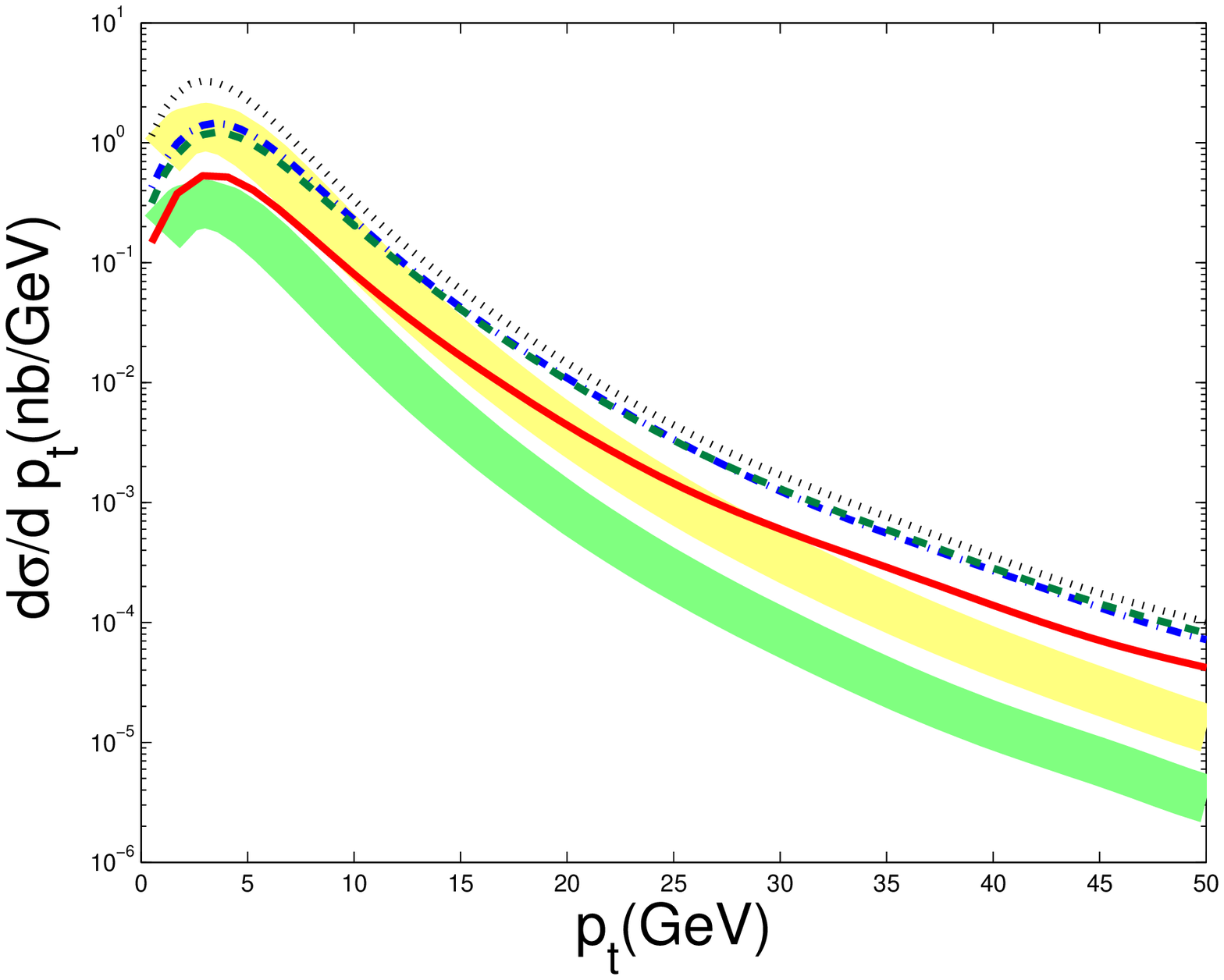}%
\hspace{5mm}
\includegraphics[width=0.460\textwidth]{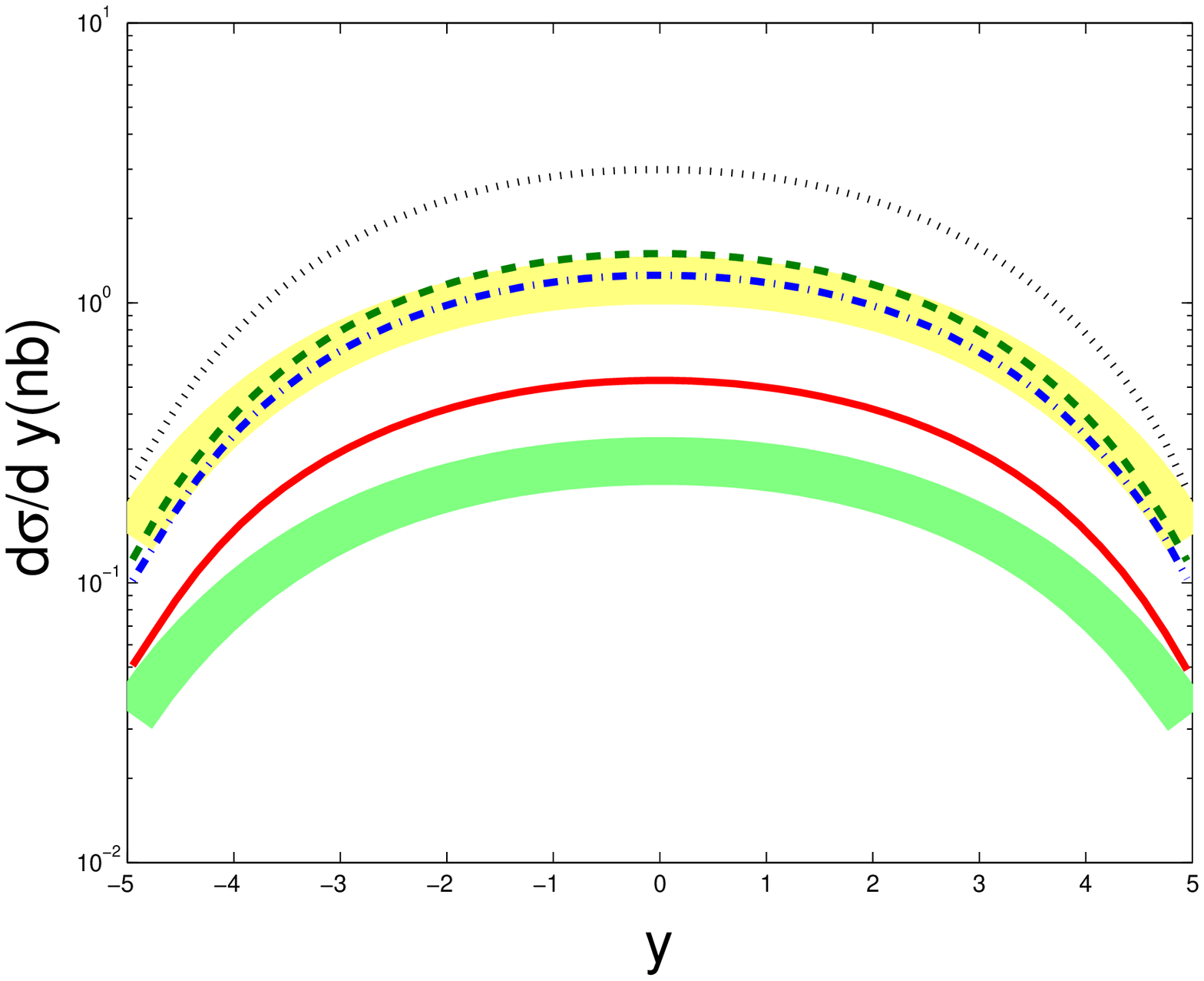}\hspace*{\fill}
\caption{\small Distributions in $p_t$ and $y$
of the hadronic production of $(c\bar{b})$ meson at LHC. The
dashed line, solid line, dash-dot line, dotted line
represent the color-singlet $^1P_1$, $^3P_0$, $^3P_1$, and $^3P_2$,
respectively. The lower and upper shaded bands stand for the color-octet
$^1S_0$ and $^3S_1$ states respectively, whose upper limit
corresponds to $\Delta_S(v)=0.3$ and lower limit
corresponds to $\Delta_S(v)=0.1$.} \label{fig1} \vspace{-0mm}
\end{figure}

\begin{figure}
\centering
\includegraphics[width=0.460\textwidth]{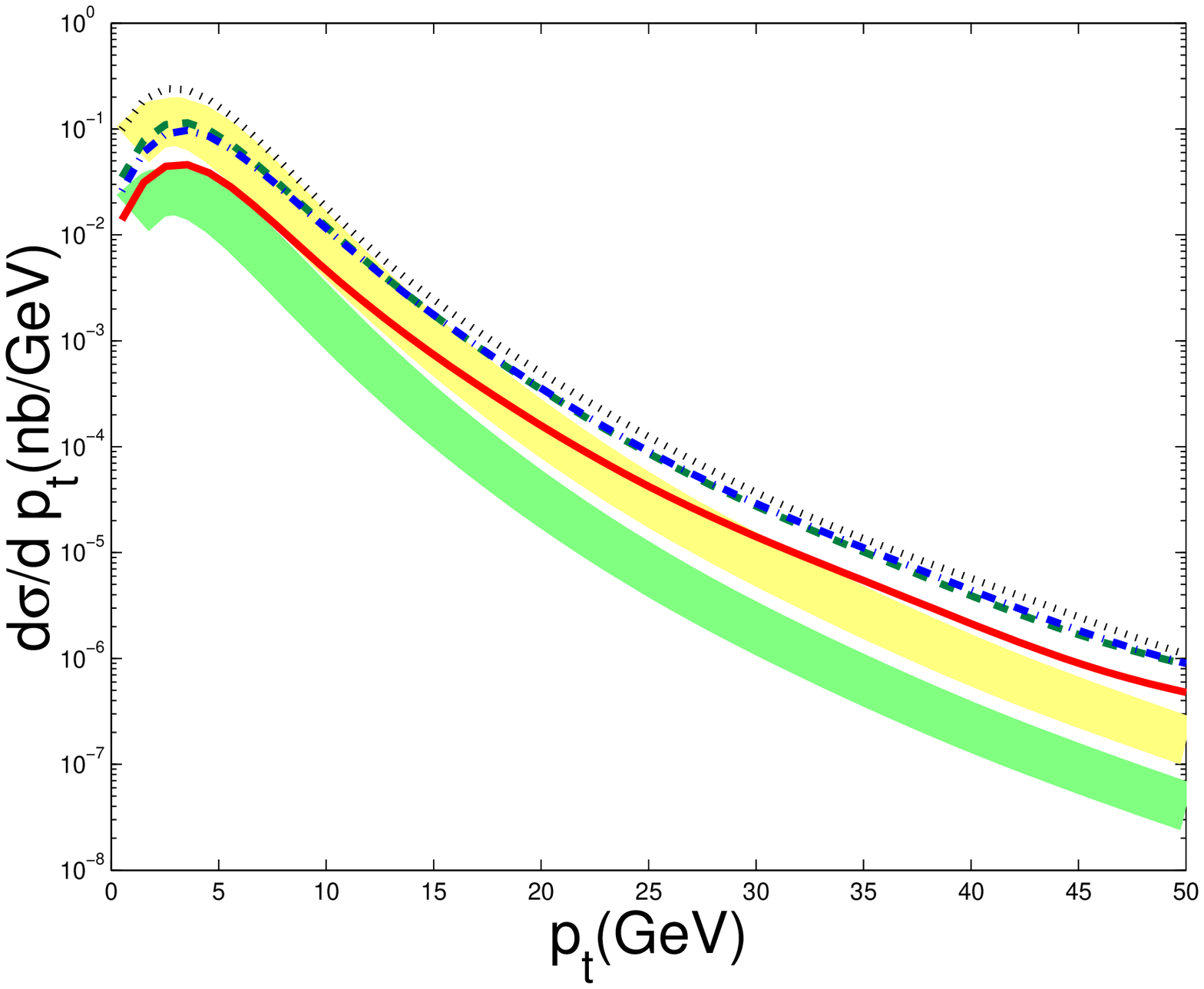}%
\hspace{5mm}
\includegraphics[width=0.460\textwidth]{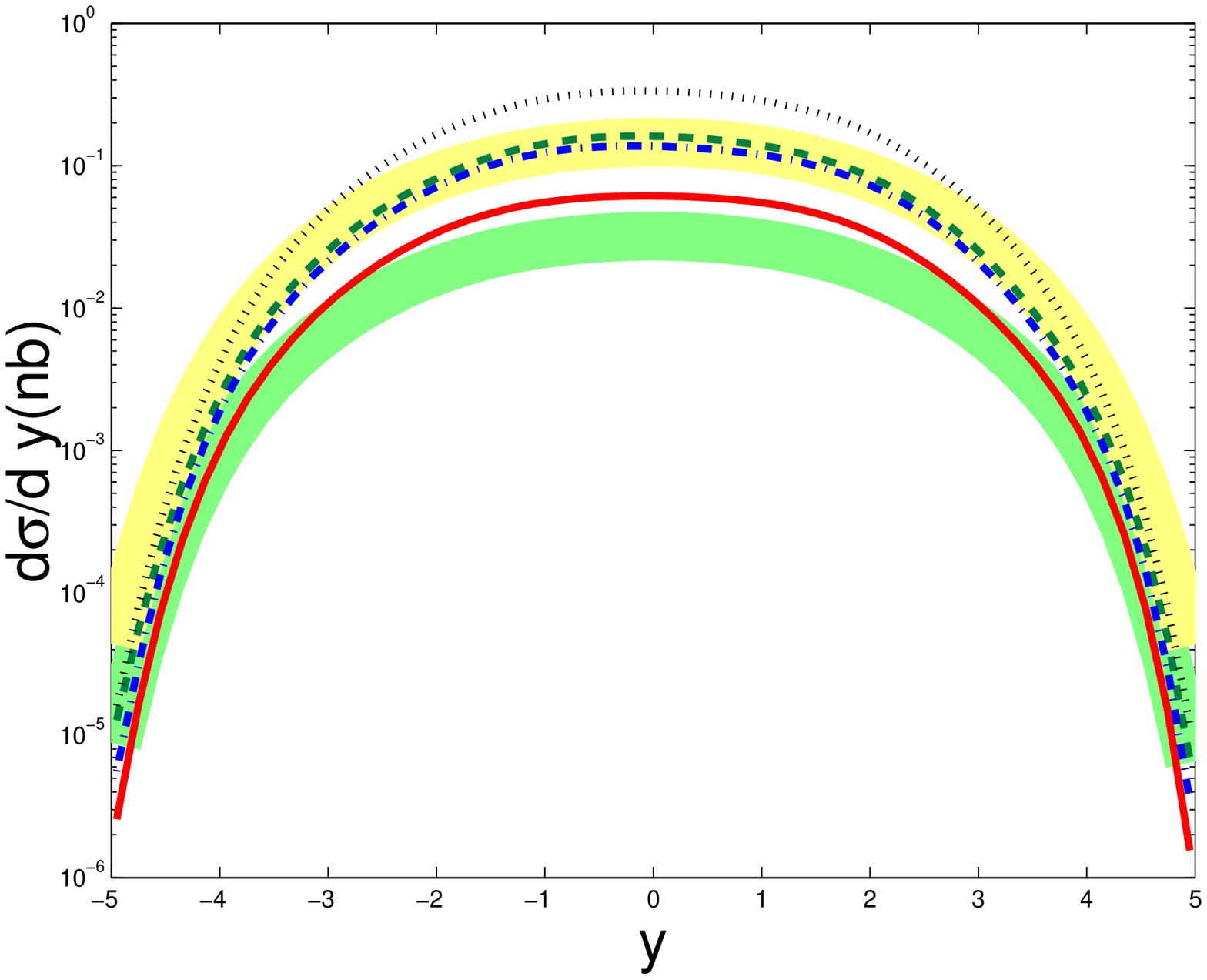}\hspace*{\fill}
\caption{\small Distributions in $p_t$ and $y$ of the hadronic
production of $(c\bar{b})$ meson at TEVATRON. The dashed line,
solid line, dash-dot line, and dotted line represent the
color-singlet $^1P_1$, $^3P_0$, $^3P_1$ and $^3P_2$, respectively.
The lower and upper shaded bands stand for color-octet $^1S_0$ and
$^3S_1$ states respectively, whose upper limit corresponds to
$\Delta_S(v)=0.3$ and lower limit corresponds to
$\Delta_S(v)=0.1$.} \label{fig2} \vspace{-0mm}
\end{figure}

To see this point more clearly, we calculate the distributions of
the corresponding bound states on the transverse momentum $p_t$
and the rapidity $y$ respectively, and plot the results in
Figs.(\ref{fig1},\ref{fig2}). From the figures Figs.(\ref{fig1},
\ref{fig2}), one may observe that the contributions to the total
cross section from the color-octet component are in comparable
with those from the color-singlet component in the $P$-wave $B_c$
meson hadronic production. However, in the $p_t$ distribution the
color-octet contributions drop more rapidly than those from the
color-singlet parts, which exhibits more clearly through looking
at the ratio, $R(p_{tcut})=\frac{\sigma_{octet}}
{\sigma_{singlet}}$, where $\sigma_{octet}$ stands for the total
contribution from the color-octet and $\sigma_{singlet}$ for that
from the color-singlet. The variation of the dependence of
$R(p_{tcut})$ on $p_{tcut}$ is presented in TABLE.\ref{ratio}.
Here, for simplicity, we take a middle value of $\Delta^2_S(v)$,
i.e. $\Delta^2_S(v)=0.05$ (or $\Delta_S(v)=0.224$), for the
color-octet matrix elements.

\begin{table}
\begin{center}
\caption{The correlation of $R(p_{tcut})=\frac{\sigma_{octet}}
{\sigma_{singlet}}$ with the $p_{tcut}$. For color-octet matrix
elements, we take $\Delta^2_S(v)=0.05$, which
is in the middle of the range $(0.01,0.09)$.} \vspace{2mm}
\begin{tabular}{|c||c|c|c|c|c|}
\hline\hline - &~$p_{tcut}$=0 GeV~ & ~$p_{tcut}$=5 GeV~ &
~$p_{tcut}$=20 GeV~ & ~$p_{tcut}$=35 GeV~& ~$p_{tcut}$=50 GeV~ \\
\hline\hline LHC & 0.22 & 0.19 & 0.095 & 0.058 & 0.043 \\
\hline TEVATRON & 0.25 & 0.20 & 0.098 & 0.060 & 0.044 \\
\hline\hline
\end{tabular}
\label{ratio}
\end{center}
\end{table}

\begin{figure}
\centering
\includegraphics[width=0.460\textwidth]{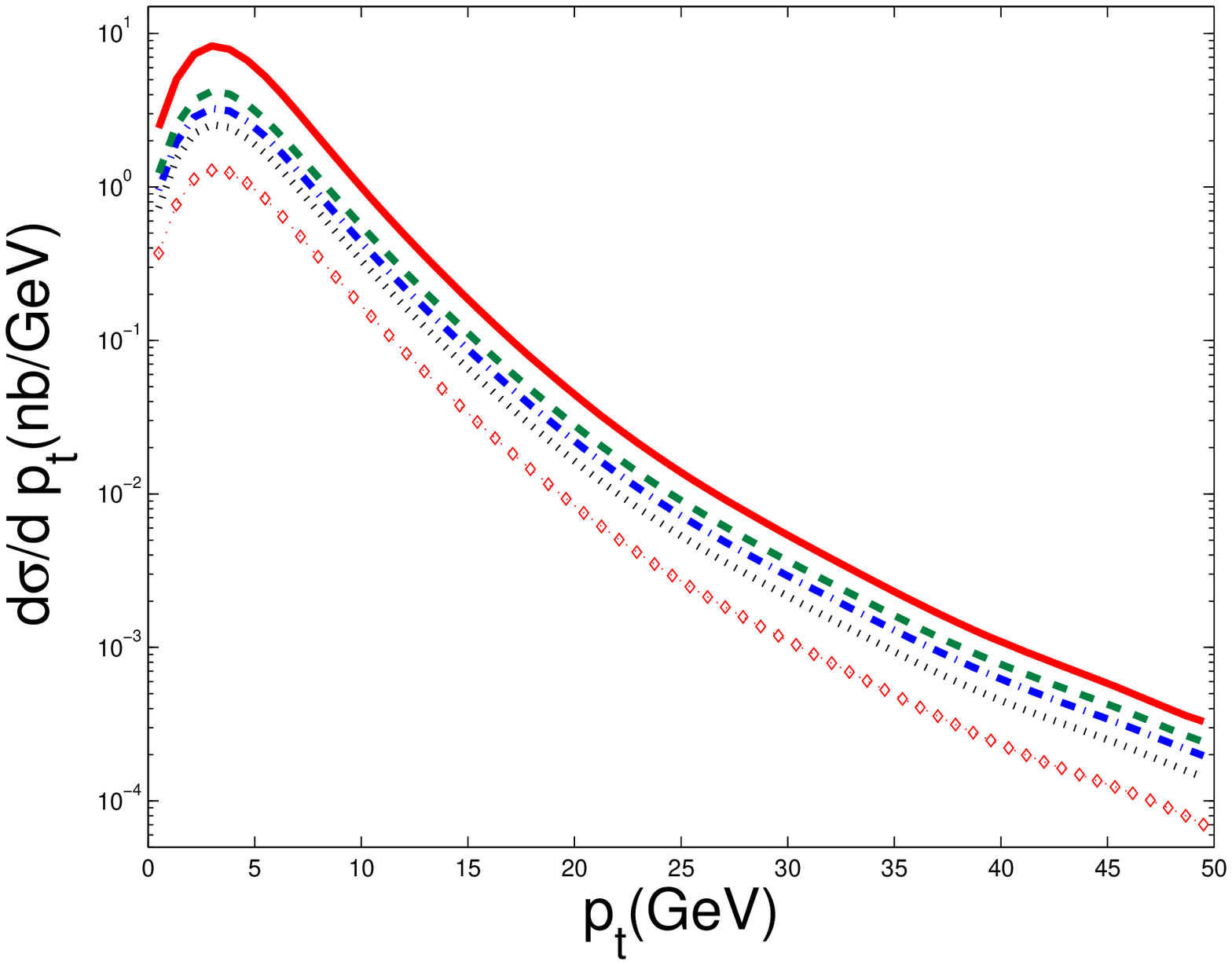}%
\hspace{5mm}
\includegraphics[width=0.460\textwidth]{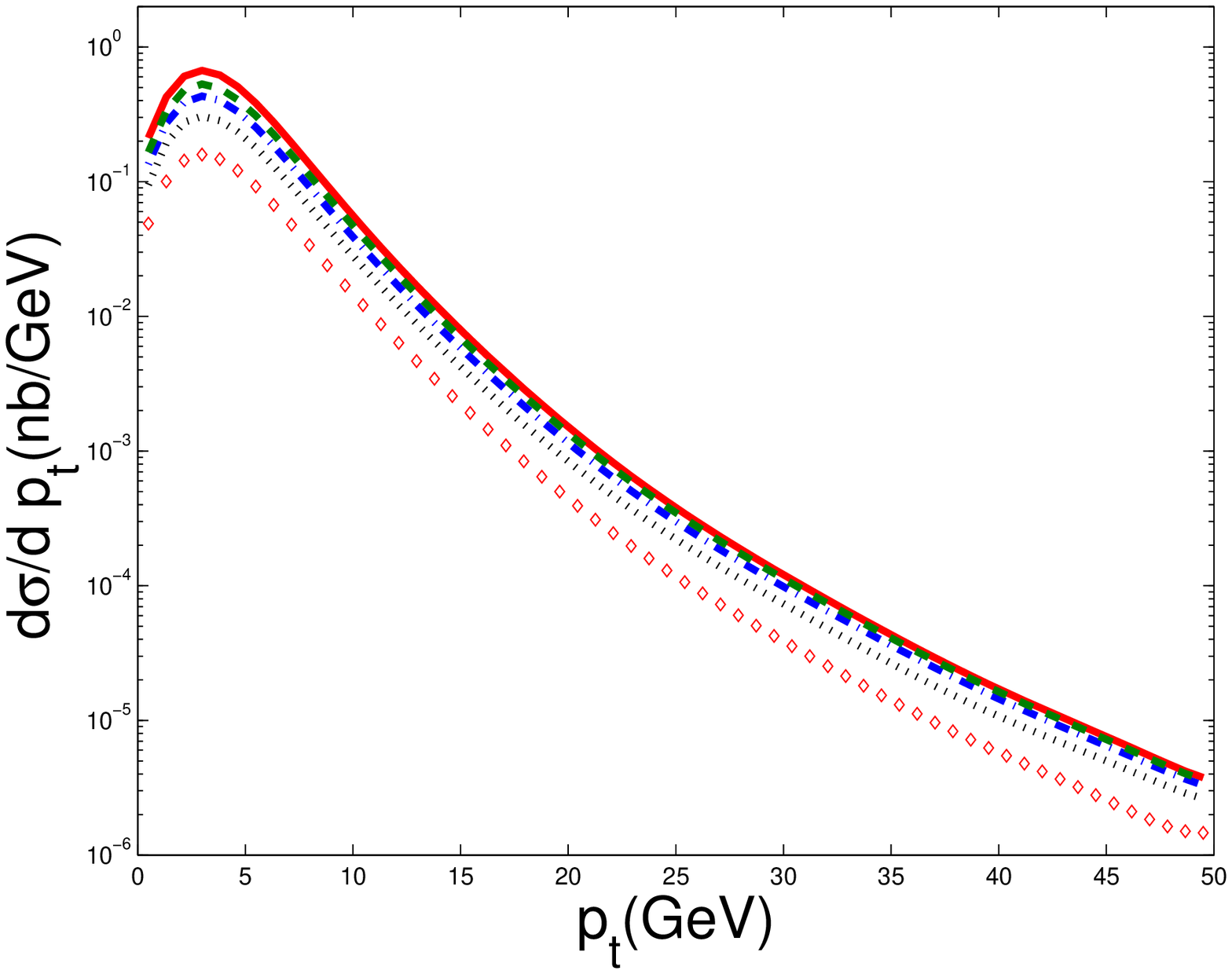}\hspace*{\fill}
\caption{\small $p_t$ distributions of various contributions in $h_{B_c}$
and $\chi_{B_c}^J$ production, with several different cuts in rapidity
($y_{cut}$) at LHC (left) and at TEVATRON (right) energies.
Dashed line (next to top) with $y_{cut}=2.0$; dash-dot
line (middle) with $y_{cut}=1.5$; dotted line (next bottom) with
$y_{cut}=1.0$; diamond line (bottom) with $y_{cut}=0.5$ and solid
line (top) without $y_{cut}$. The color-octet matrix elements are
set at $\Delta^2_S(v)=0.05$.} \label{lypt} \vspace{-0mm}
\end{figure}

\begin{figure}
\centering
\includegraphics[width=0.460\textwidth]{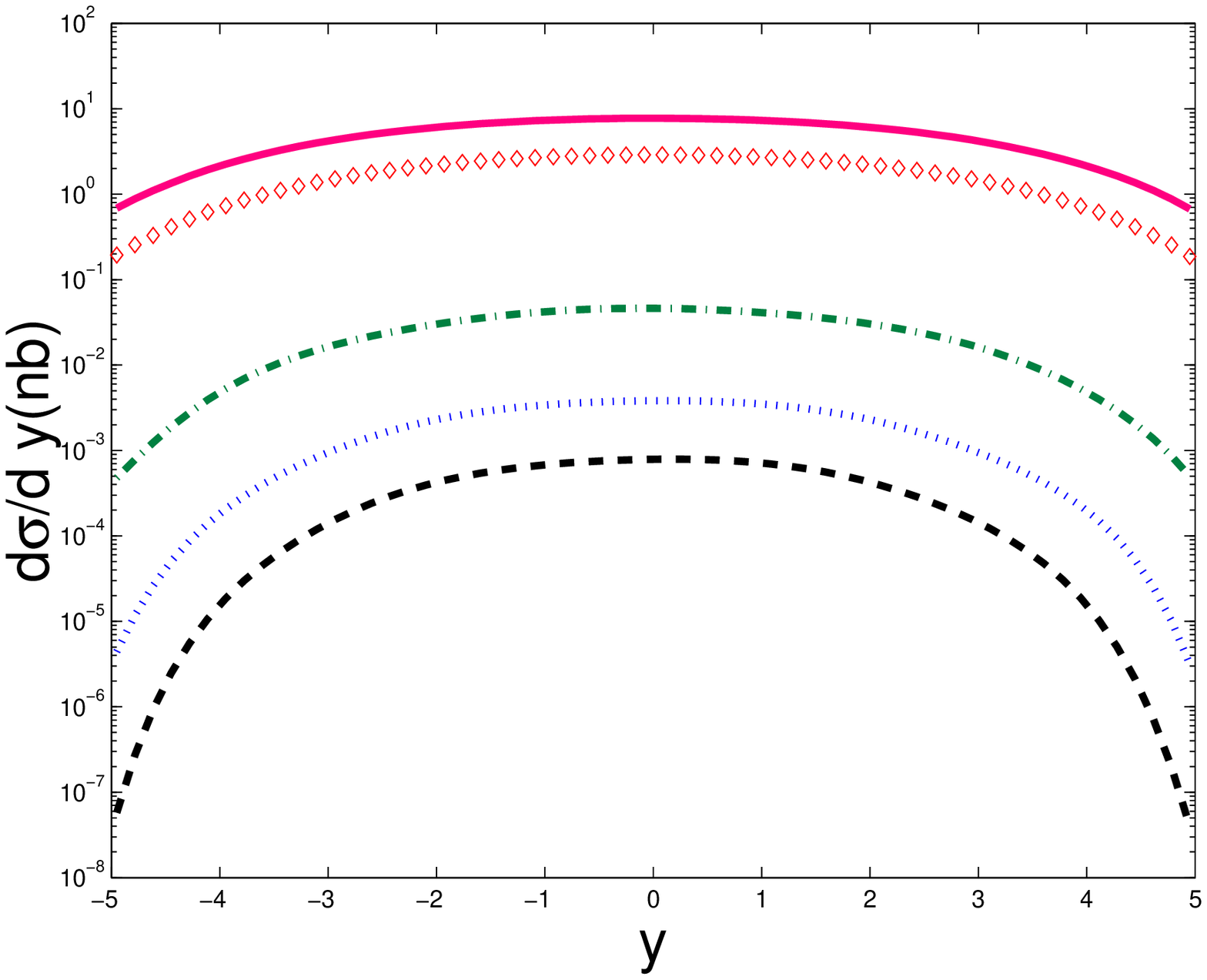}%
\hspace{5mm}
\includegraphics[width=0.460\textwidth]{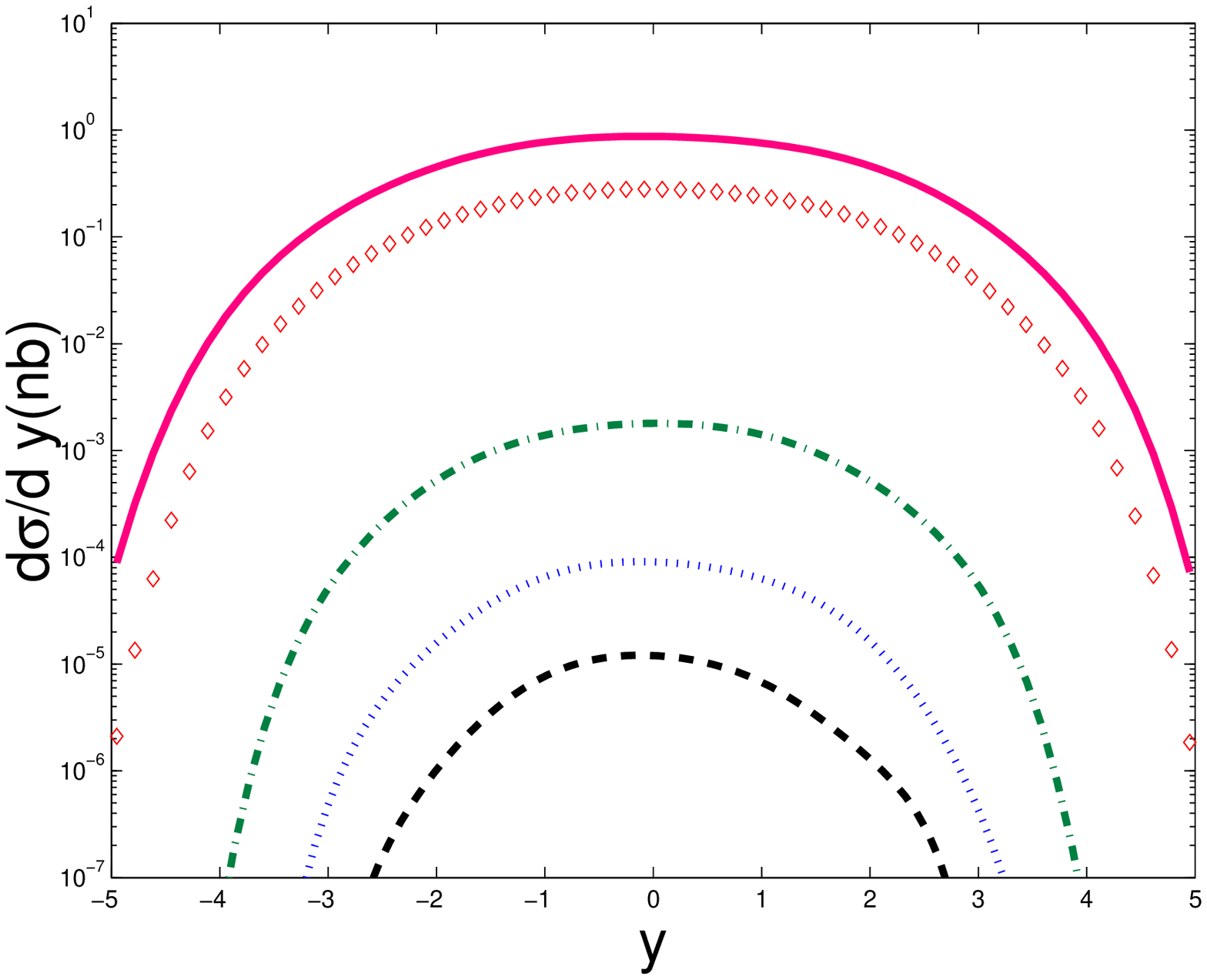}\hspace*{\fill}
\caption{\small $y$ distributions of all the considered
contributions, i.e., from both color-octet $S$-wave states and
color-singlet $P$-wave states, with various transverse momentum
cuts ($y_{cut}$) at LHC (left) and at TEVATRON (right) energies.
Here, the Diamond line represents what with $p_{tcut}=5$ GeV;
dash-dot line with $p_{tcut}=20$ GeV; dotted line with
$p_{tcut}=35$ GeV; dashed line with $p_{tcut}=50$ GeV, and the
solid line with $p_{tcut}=0$. Of the color-octet matrix elements,
we take $\Delta^2_S(v)=0.05$.}
\label{lpty} \vspace{-0mm}
\end{figure}

Since there is no detector in high energy hadronic collisions
which can directly detect all the events, especially those with a
small $p_t$ and/or a large rapidity $y$, so for experimental
studies and for practical purpose, only the events with proper
kinematic cuts on $p_t$ and/or $y$ are taken into account. Now let
us compute the observables with cuts on $p_t$ and/or $y$, so as to
see the the cut effects in $h_{B_c}$ and $\chi_{B_c}^J$ production
precisely. Namely we have investigated the effects in production
rate and various differential cross sections and for the
color-octet production and the color-singlet production of
$S$-wave ($B_c$ and $B_c^*$) and $P$-wave ($h_{B_c}$ and
$\chi_{B_c}^J$) respectively. Explicitly, we take
$\Delta^2_S(v)=0.05$ for the color-octet matrix elements in the
investigation. Considering the abilities on measuring $p_t$ and
$y$ rapidity of $B_c$ for CDF, D0, BTeV at TEVATRON and for ATLAS,
CMS, and LHC-B at the LHC, we compute the $p_t$ distributions with
the rapidity cuts $y_{cut}= 1.5$, and the $y$ distributions with
the transverse momentum cut $p_{tcut}= 5$ GeV accordingly. The
results with four rapidity cuts, $y_{cut}=(0.5, 1.0, 1.5, 2.0)$,
are put together in Fig.(\ref{lypt}) and the results with four
transverse momentum cuts, $p_{tcut}=5, 20, 35, 50$ GeV, are put in
Fig.(\ref{lpty}).

\section{Discussions and summary}

In the paper, according to the NRQCD expectation about the
importance for the components, we have precisely investigated the
contributions of the $S$-wave color-octet components to the
hadronic production of the $P$-wave $B_c$ states. Our final
results show that contributions from the color-octet ones are
comparable to those from the color-singlet ones in the production
of the $P$-wave $B_c$ excited states if the scale rule of NRQCD
works well and the value of the relative velocity squared $v^2$ is
really in the possible region $0.1\sim 0.3$, as indicated by
potential models for the $(c\bar b)$ binding system for instance.
Therefore, to make a soundly prediction to the hadronic production
of the $P$-wave $(c\bar{b})$ meson at leading order in $v^2$, one
needs to take both contributions from the color-octet $S$-wave
Fock states and those from the color-singlet $P$-wave Fock states
into account. With $\Delta^2_S(v)=0.05$ and $p_{tcut}=0.0$ GeV,
the total contributions from the color-octet $S$-wave Fock states
to the total production cross section is about $\sim 22\%$ from
the color-singlet $P$-wave states at LHC and about $\sim 25\%$ at
TEVATRON respectively, and such contributions decrease with the
increment in $p_{tcut}$, e.g. at $p_{tcut}=50$ GeV, it reduces to
$\sim 4\%$ at both energies of LHC and TEVATRON. This character
may used for people to distinguish the color-singlet contributions
from those from color-octet ones in future when there are enough
$h_{B_c}$ and $\chi_{B_c}^J$ data.

Note that for shortening the paper and not changing the main
feature, here we have not taken the possible mixing between
$h_{B_c}$ and $\chi^{J=1}_{B_c}$ into account at all. According to
NRQCD scale rule, except the components $|(c\bar b)_{\bf
8}(^{1}S_{0})g\rangle)$ and $|(c\bar b)_{\bf
8}(^{3}S_{1})g\rangle)$, there is no enhancement factor, such as
that comes from the wave function at the origin for the $S$-wave
components virus the derivative of the wave function at the origin
for the leading $P$-wave color-singlet ones, so we consider the
contributions from the other high order Fock components in
Eq.(\ref{eq:2}) small, and ignore them at all.

In summary, the total cross-section of the $h_{B_c}$ and
$\chi_{B_c}^J$ production, including both singlet and octet
contributions, at the lowest order in $\alpha_s$ and $v^2$
expansions, may be so bigger than a half of the direct production
rate of the ground state $B_c(^1S_0)$ (see TABLE.II, roughly
speaking: $\sim 0.8$ for LHC and $\sim 0.7$ for TEVATRON).
Considering the fact that almost all of the low-laying excited
states decay to the ground state $B_c$, it also means that the
full low-laying $P$-wave state production `promptly' contributes
the $B_c(^1S_0)$ production by a bigger factor than $0.5$ of the
direct one. Especially, if $\Delta^2_S(v)\sim 0.05$, the
contributions from color-octet components themselves may
contribute promptly a factor $\sim 0.1$ of the direct hadronic
production of $B_c$ meson. Furthermore, our complete calculations
also indicate that the observation of the $P$-wave $B_c$ states is
not very far from the present experimental capability in
principle.

\vspace{20mm} \noindent {\Large\bf Acknowledgements:} We are
grateful to G.T. Bodwin for drawing our attention to the research
topic of the paper. This work was supported in part by the Natural
Science Foundation of China (NSFC).\\

\end{document}